\documentclass{llncs}
\usepackage{graphicx}
\usepackage{graphics}
\usepackage{epsfig}
\usepackage{epstopdf}
\usepackage[cmex10]{amsmath}
\usepackage{algorithmic}
\usepackage{array}
\usepackage{mdwmath}
\usepackage{mdwtab}
\begin{document}

\title{A Synthesis Method for\\ Quaternary Quantum Logic Circuits}

\author{Sudhindu Bikash Mandal\inst{1}
\hspace{0.2in} Amlan Chakrabarti$^1$ 
\hspace{0.2in} Susmita Sur-Kolay\inst{2}
}

\institute{A. K. Choudhury School of Information Technology, University of Calcutta, India
\and
Advanced Computing and Microelectronics Unit, Indian Statistical Institute, India
}

\maketitle
\begin{abstract}
Synthesis of quaternary quantum circuits involves basic quaternary gates and logic operations in the quaternary quantum domain. In this paper, we propose new projection operations and quaternary logic gates for synthesizing quaternary logic functions. We also demonstrate the realization of the proposed gates using basic quantum quaternary operations. We then employ our synthesis method to design of quaternary adder and some benchmark circuits. Our results in terms of circuit cost, are better than the existing works.

\end{abstract}
\vspace*{-5mm}
\section{Introduction}
\vspace*{-3mm}
Quaternary quantum computing is gaining importance in the field of quantum information theory and quantum cryptography as it can represent a Galois Field(4) quantum system by the basis states $|0\rangle$, $|1\rangle$, $|2\rangle$ and $|3\rangle$. The unit of information is called a {\it qudit} which is characterized by a wave function $|\psi\rangle$ \cite{Nielsen,Ekert} expressed as a linear superposition of basis states. Multi-valued quantum algebra comprises the rules for a set of basic logic operations that can be performed on qudits. While in \cite{Muthukrishnan} the structure of a multi-valued logic gate is proposed which is experimentally feasible with a linear ion trap scheme for quantum computing; this approach can produce large dimensional circuits. A universal architecture for multi-valued reversible logic is given in \cite{Klimov}, but quantum realization of the circuits thus obtained is not apparent. The universality of $n$-qudit gates is presented in \cite{Zilic}, but no algorithms for synthesis were given. Al-Rabedi et al. proposed in \cite{Yang} the minimization technique for multi-valued quantum Galois field sum of products (QGFSOP). Quaternary logic is one of the promising multi-valued quantum logic systems. The binary logic functions can be expressed by grouping $2$-bits together into equivalent quaternary value \cite{Chowdhury}. This theoretically reduces the total volume of the physical devices needed to approximately $1/log_2^4$, i.e. $1/2$ the volume needed for binary system \cite{Chowdhury}. 


The realization of a given quaternary quantum function as a quantum circuit requires a set of gates for the quaternary logic operations. In \cite{Khan}, the realization of quaternary Feynman and Toffoli gates using $1$-qudit and $2$-qudit quaternary Muthukrishnan-Stroud gates (M-S Gate)\cite{Muthukrishnan} are illustrated. The QGFSOP expressions can be realized by using these Feynman and Toffoli gates \cite{Khan}. The method of synthesizing incompletely specified multi-output quaternary function using quaternary $1$-qudit gates and multi-qudit controlled gates has been proposed in \cite{Mk}. But this synthesis method is not applicable for any arbitrary quaternary functions. In  \cite{siddi}, a heuristic alogorithm is proposed for minimization of a QGFSOP expression for multi-output quaternary logic functions using a Quaternary Galois Field Decission Diagram (QGFDD). But no quantum gate level implementation was provided. In this paper, our specific contributions are as follows:
\begin{itemize}
\item[$\bullet$] new projection operation for synthesizing quaternary logic functions;
\item[$\bullet$] new  quaternary logic gates, namely Generalized Quaternary Gate ($GQG$), permutative quaternary Controlled Cyclic Shift gate ($C^2CS$) and Modulo$4$ addition gate;
\item[$\bullet$] new simplification rules for reduction in gate count and circuit levels for multivalued quantum circuits.
\end{itemize}
\par
The rest of the paper is organized as follows. We provide the preliminary concepts of multivalued quantum computing in section 1. We propose the qua-
ternary algebra with a new projection operation in section 2. In section 3, we introduce some new quaternary logic gates. The proposed synthesis methodology along with its simplification rules are presented in section 4. The synthesis results for some example circuits and their comparison with related work are given in section 5. Concluding remarks appear in section 6.
\vspace*{-5mm}
\section{Quaternary Algebra}
\vspace*{-3mm}
A brief summary of the quaternary addition, multiplication and NOT operations as well as the quaternary projection operations $L$, $J$ and (the new one) $P$ are presented next. .
\vspace*{-5mm}
\subsection{GF(4) arithmetic}
Quaternary Galois field GF(4) is an algebraic structure consisting the set of elements Q=\{0, 1, 2, 3\}. The addition ($+$) and multiplication ($.$) operations over GF(4) are shown in the Table 1.
\vspace*{-7mm}
\begin{table}
\caption{GF(4) Addition and Multiplication }
\vspace*{-7mm}
\begin{center}
\begin{tabular}{|r|r|r|r|r||r|r|r|r|r|}
\hline
$+$ & $0$& $1$& $2$& $3$ & $.$ & $0$& $1$& $2$& $3$\\
\hline
0 & 0 & 1 & 2 & 3 & 0 & 0 & 0 & 0 & 0\\
1 & 1 & 0 & 3 & 2 & 1 & 0 & 1 & 2 & 3\\
2 & 2 & 3 & 0 & 1 & 2 & 0 & 2 & 3 & 1\\
3 & 3 & 2 & 1 & 0 & 3 & 0 & 3 & 1 & 2\\
\hline
\end{tabular}
\end{center}
\end{table}
\vspace*{-15mm}
\subsection{Quaternary Logical NOT}
\vspace*{-3mm}
The logical NOT in quaternary quantum system is defined as
$NOT(a)=a+1$, where $'+'$ denotes the modulo 4 addition, and $a$ =\{0, 1, 2, 3\}.
\vspace*{-5mm}
\subsection{Quaternary Projection Operations $L$, $J$ and $P$}
\vspace*{-3mm}
We present nine projection operations, grouped into three types $L_i$, $J_i$, and $P_i$, where $i$ = \{$0$, $1$, $2$, $3$\}. While $L_i$ and $J_i$ types were defined earlier \cite{SKolay,Herman} as \\
\indent  $L_i(a)$ = $1$ if $a$ = $i$ and $0$ otherwise, and \\
\indent $J_i(a)$ = $2$ if $a$ = $i$ and $0$ otherwise. \\
We introduce the new $P_i$ type operations, which are defined as \\
\indent $P_i(a)$ = 3 if a = $i$ and a = $0$ otherwise. \\
Table 2 presents the truth tables for $L_i$, $J_i$, and $P_i$ types of operators as well as for the derived operators  $L^\prime_{i}$, $J^\prime_{i}$ and $P^\prime_{i}$. The $L_i$, $J_i$ and $P_i$ operations are commutative, associative and distributive over AND and OR logic.
\begin{table}
\caption{Truth table of projection operations $L_i$, $L^\prime_{i}$, $J_i$, $J^\prime_{i}$, $P_i$, $P^\prime_{i}$}
\begin{center}
\vspace*{-5pt}
\begin{tabular}{|c|c|c|c|c||c|c|c|c|c||c|c|c|c|c||c|c|c|c|c|}
\hline
$a$ & 0 & 1 & 2 & 3 & $a$ & 0 & 1 & 2 & 3 & $a$ & 0 & 1 & 2 & 3 & $a$ & 0 & 1 & 2 & 3\\
\hline
$L_0(a)$ & 1 & 0 & 0 & 0 & $J_2(a)$ & 0 & 0 & 2 & 0 & $L^\prime_{0}(a)$ & 0 & 1 & 1 & 1 & $J^\prime_{2}(a)$ & 2 & 2 & 0 & 2\\
\hline
$L_1(a)$ & 0 & 1 & 0 & 0 & $J_3(a)$ & 0 & 0 & 0 & 2 & $L^\prime_{1}(a)$ & 1 & 0 & 1 & 1 & $J^\prime_{3}(a)$ & 2 & 2 & 2 & 0\\
\hline
$L_2(a)$ & 0 & 0 & 1 & 0 & $P_0(a)$ & 3 & 0 & 0 & 0 & $L^\prime_{2}(a)$ & 1 & 1 & 0 & 1 & $P^\prime_{0}(a)$ & 0 & 3 & 3 & 3\\
\hline
$L_3(a)$ & 0 & 0 & 0 & 1 & $P_1(a)$ & 0 & 3 & 0 & 0 & $L^\prime_{3}(a)$ & 1 & 1 & 1 & 0 & $P^\prime_{1}(a)$ & 3 & 0 & 3 & 3\\
\hline
$J_0(a)$ & 2 & 0 & 0 & 0 & $P_2(a)$ & 0 & 0 & 3 & 0 & $J^\prime_{0}(a)$ & 0 & 2 & 2 & 2 & $P^\prime_{2}(a)$ & 3 & 3 & 0 & 3\\
\hline
$J_1(a)$ & 0 & 2 & 0 & 0 & $P_3(a)$ & 0 & 0 & 0 & 3 & $J^\prime_{1}(a)$ & 2 & 0 & 2 & 2 & $P^\prime_{3}(a)$ & 3 & 3 & 3 & 0\\
\hline
\end{tabular}
\end{center}
\end{table}
\vspace*{-15mm}
\section{Quaternary Logic Gates}
\vspace*{-2mm}
The definitions of the existing quaternary logic gates as well as a few newly introduced quaternary logic gates are provided below. We also show the implementation of the newly proposed gates using basic quantum ternary operations.
\vspace*{-7mm}
\subsection{Quaternary Feynman, Quaternary Toffoli, MAX and MIN gates}
\vspace*{-2mm}
The $2$-qudit Quaternary Feynman gate \cite{Khan} is defined as: \\
\indent Feynman($A$, $B$) = $A$ + $B$, where '+' operator is addition over GF(4)(Figure 1.a). \\
The quaternary $3$-qudit Toffoli gate \cite{Khan}, shown in Figure 1.b is defined as: \\
\indent Toffoli$(A, B, C)=A.B+C$, where $A$ and $B$ are the control inputs and $C$ the target input, '+' and '.' operators are addition and multiplication over GF(4). We use the quaternary $MAX$ and $MIN$ gates \cite{Gieseck} respectively to replace the OR and the AND gates. These two gates are defined as \\
MAX($A_1,A_2,..,A_n$, $B$)=$\left\{\begin{array}{ll}\mbox{$A_i$} & \mbox{if $A_i$ $\geq$ $A_j$, $i \neq j$ and  $A_i$ $\geq$ $B$};\\
                                                         \mbox{$B$} &  \mbox{if $\forall i, B \geq A_i$};\end{array}\right.$\\
MIN($A_1,A_2,...A_n$, $B$) =$\left\{\begin{array}{ll}\mbox {$A_i$} & \mbox{ if $A_i$ $\leq$ $A_j$, $i \neq j$ and $A_i \leq B$};\\
							 \mbox {$B$} & \mbox{if $\forall i, B \leq A_i$};\end{array}\right.$\\
where $i$ = $\{1,2,3...n\}$ (Figures 2.a, 2.b). The quaternary Feynman and Toffoli gates can be realized using quaternary M-S gates \cite{Chowdhury,Khan}, defined as (Figure 2.c)\\ 
\indent M-S($A$, $B$) = $Z$ operation on $B$ if $A$ = $3$, otherwise $B$, where $Z$ is one of the $24$ shift operations \cite{siddi} shown in Table 3.
\vspace*{-7mm}
\begin{table}
\caption{ Shift Operations over GF(4) [10]}
\begin{center}
\begin{tabular}{|r|r||r|r||r|r||r|r|}
\hline
Symbol & Operation & Symbol & Operation & Symbol & Operation & Symbol & Operation\\
\hline
$x^{+0}$ & $x$ = $x$ &$x^{021}$ & $x$ = $2x+2$ & $x^{23}$ & $x$ = $x^2$& $x^{0231}$ & $x$ = $2x^2+2$\\
$x^{+1}$  & $x$ = $x+1$ &$x^{032}$ & $x$ = $2x+3$ & $x^{01}$ & $x$ = $x^2+1$& $x^{03}$ & $x$ = $2x^2+3$\\
$x^{+2}$ & $x$ = $x+2$ &$x^{132}$ & $x$ = $3x$ & $x^{0213}$ & $x$ = $x^2+2$& $x^{13}$ & $x$ = $3x^2$\\
$x^{+3}$ & $x$ = $x+3$ &$x^{012}$ & $x$ = $3x+1$ & $x^{0312}$ & $x$ = $x^2+3$& $x^{0123}$ & $x$ = $3x^2+1$\\
$x^{123}$ & $x$ = $2x$ &$x^{023}$ & $x$ = $3x+2$ & $x^{12}$ & $x$ = $2x^2$& $x^{02}$ & $x$ = $3x^2+2$\\
$x^{013}$ & $x$ = $2x+1$ &$x^{031}$ & $x$ = $3x+3$ & $x^{0132}$ & $x$ = $2x^2+1$& $x^{0321}$ & $x$ = $3x^2+3$\\
\hline
\end{tabular}
\end{center}
\end{table}
\begin{figure}[!h]
\centering
\vspace*{-3mm}
\includegraphics {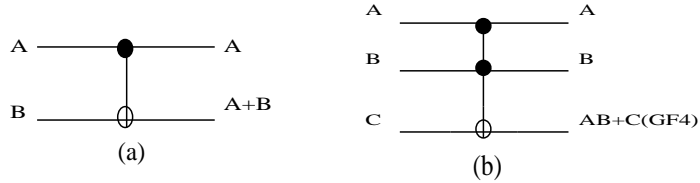}
\vspace*{-4mm}
\caption{(a) $2$-qudit Quaternary Feynman gate (b) 3-qudit quaternary Toffoli gate}
\label{Figure 1:}
\end{figure}
\begin{figure}[!h]
\centering
\vspace*{-4mm}
\includegraphics {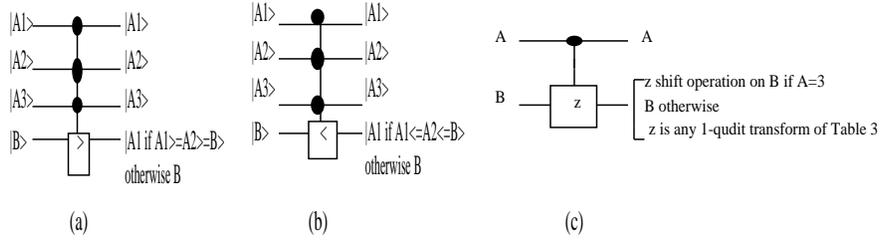}
\vspace*{-4mm}
\caption{(a) Quaternary MAX gate, (b) Quaternary MIN gate, (c) Quaternary M-S gate}
\label{Figure 2:}
\end{figure}
\vspace*{2mm}
\subsection{Generalized Quaternary Gate}
A new generalized quaternary gate (GQG) is required to realize the 24 shift operations \cite{siddi} given in the Table 3. It is a multi-qudit gate shown in Figure 3.a. The controlling input of GQG can be used to select the $1$-qudit shift operation on the target input. The GQG is formally defined as\\
$GQG(A_1, A_2,..,A_n, B)$=$\left\{\begin{array}{llll}\mbox{$B$ shift $X$} & \mbox{if $A_1, A_2,..,A_n = 0$};\\
                                                         \mbox{$B$ shift $Y$} &  \mbox{if $A_1, A_2,..,A_n = 1$};\\
                                                         \mbox{$B$ shift $Z$} &  \mbox{if $ A_1, A_2,..,A_n = 2$};\\ 
\mbox{$B$ shift $W$} &  \mbox{if $ A_1, A_2,..,A_n = 3$};\\
				         \mbox{$B$ } &  \mbox{otherwise};\end{array}\right.$\\
The realization of a GQG gate using quaternary M-S gates, is shown in Figure 3.b.
\begin{figure}[!h]
\centering
\vspace*{-9mm}
\includegraphics {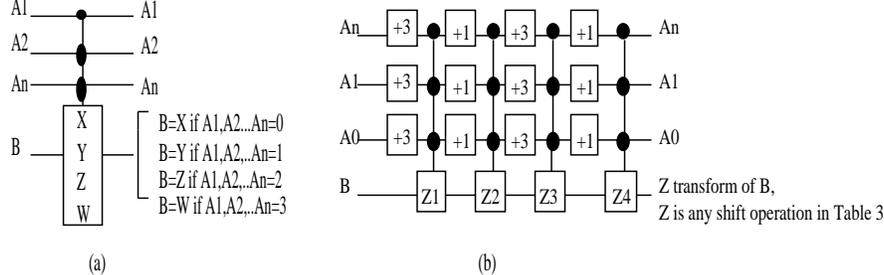}
\vspace*{-5mm}
\caption{(a) A multi-qudit generalized quaternary gate (GQG), and (b) its realization using M-S gates}
\label{Figure 5:}
\end{figure}
\vspace*{-7mm}
\subsubsection{Implementation of $L_i$, $J_i$ and $P_i$ operations using GQG}
\vspace*{-3mm}
We can implement the $L_i$, $J_i$ and $P_i$ operations by using a GQG, as shown in Figures 4.a, 4.b and 4.c respectively. For $L_i$, $J_i$ and $P_i$ type operations, we set GQG(a,1), GQG(a,2) and GQG(a,3) respectively.\\
\begin{figure}[!h]
\centering
\vspace*{-10mm}
\includegraphics {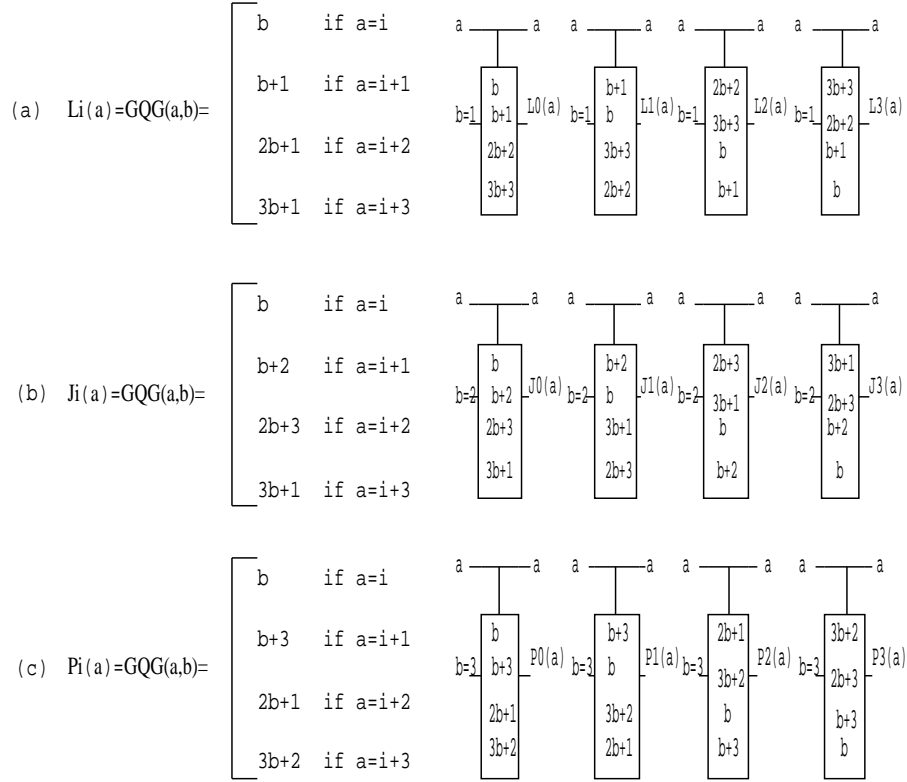}
\caption{GQG based realization of quaternary projection operations: (a)$L_i$, (b)$J_i$ and (c)$P_i$}
\label{Figure 3a:}
\end{figure}
\vspace*{-7mm}
\subsection{Quaternary Controlled Cyclic Shift $C^2CS$ gate}
\vspace*{-3mm}
We propose a new $3$-qudit $C^2CS$ gate, used for realizing the simplification rules for quaternary minterms, as\\
$C^2CS(A,B,C)=\left\{\begin{array}{ll}\mbox{$C^{0123}$} & \mbox{if $A\neq B$};\\
                                                         \mbox{$C$} &  \mbox{otherwise};\\
                                                          \end{array}\right. $
where the values of the inputs are from the set \{1, 2, 3\} and $C^{0123}$ is the cyclic shift operation $x^{0123}$ defined in Table 3. The symbolic representation of $C^2CS$ is shown in Figure 5.a and an instance of this gate is shown in Figure 5.b, where $x^{0123}$ shift operation is applied on $C$ if $A=1$, $B=3$ or $A=3$, $B=1$. The realization of the $C^2CS$ gate using M-S gates is shown in Figure 6.
\vspace*{-5mm}
\begin{figure}[!h]
\centering
\vspace*{-1mm}
\includegraphics {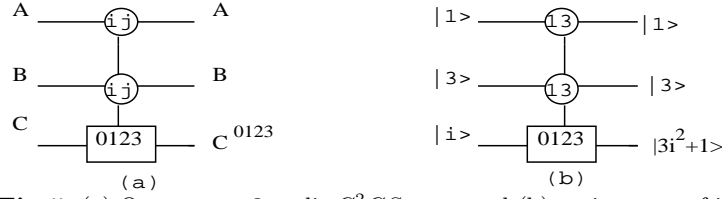}
\vspace*{-5mm}
\caption{(a) Quaternary $3$-qudit $C^2CS$ gate, and (b) an instance of it. }
\vspace*{-2mm}
\label{Figure 4:}
\end{figure}
\begin{figure}[!h]
\centering
\includegraphics {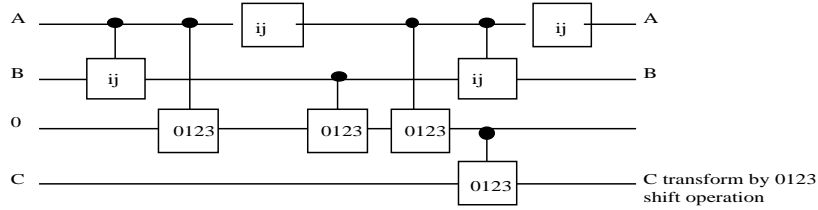}
\vspace*{-2mm}
\caption{Realization of quaternary 3-qudit $C^2CS$ gate with M-S gates}
\label{Figure 2:}
\end{figure}
\subsection{A new Modulo$4$ Addition Gate}
We introduce the new $2$-qudit quaternary Modulo$4$ Addition gate as: ADD(A, B)= $A$ $\bigoplus$ $B$, where $\bigoplus$ represent the modulo $4$ addition. This gate can be used as a template for simplification of adder circuits. The symbolic representation of ADD gate is shown in Figure 7.a and the realization using quaternary M-S gate is shown in Figure 7.b
\begin{figure}[!h]
\centering
\vspace*{-4mm}
\includegraphics {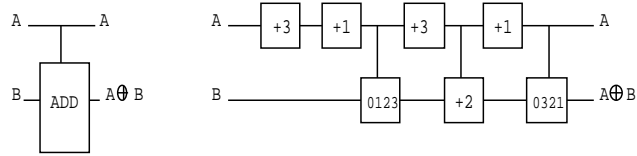}
\vspace*{-2mm}
\caption{(a) $2$-qudit Modulo$4$ Addition Gate, and (b) its realization with M-S gates}
\label{Figure 2:}
\end{figure}
\vspace*{-10mm}
\section{Proposed Synthesis Methodology}
\vspace*{-5mm}
\subsection{Overview}
\vspace*{-3mm}
Consider an $m$-variable quaternary quantum logic function\\
\indent  $f(a_1, a_2, \ldots ,a_m)$ = $\sum^n_{i=0}$(minterms for \emph{one} )$_i$ + $ \sum^p_{j=0}$(minterms for \emph{two})$_j$ +  $ \sum^s_{k=0}$(minterms for \emph{three})$_k$,\\ 
where $\sum$ implies logical quaternary \emph{OR}, \emph{ n}, \emph{p} and \emph{s} are respectively the number of input vectors for which $f$ is $1$, $2$ and  $3$. Thus, $f$ is 0 for $(4^m-n-p-s)$ of the input vectors. We express the minterms for \emph{one}, \emph{two} and \emph{three} using the $L_i$, $J_i$ and $P_i$ operations respectively. From Table II, we can state that\\
$\prod^m_{i=1}L_0(a_i)$ = $\left\{\begin{array}{ll}\mbox{1} & \mbox{\em if $\forall i$ $a_i$ = 0};\\  
\mbox{0} & \mbox{\em if $\exists$ i $a_i$ = 1, 2 or 3};\end{array}\right.$ 
 $\prod^m_{i=1}L_1(a_i)$ = $\left\{\begin{array}{ll}\mbox{1} & \mbox{\em if $\forall i$ $a_i$ = 1};\\
\mbox{0} & \mbox{\em if $\exists$ i $a_i$ = 0, 2 or 3};\end{array}\right.$ \\
$\prod^m_{i=1}L_2(a_i)$ =$\left\{\begin{array}{ll}\mbox{1} & \mbox{\em if $\forall i$ $a_i$ = 2};\\
\mbox{0} & \mbox{\em if $\exists$ i $a_i$ = 0, 1 or 3};\end{array}\right.$  
$\prod^m_{i=1}L_3(a_i)$ =$\left\{\begin{array}{ll}\mbox{1} & \mbox{\em if $\forall i$ $a_i$ = 3};\\
\mbox{0} & \mbox{\em if $\exists$ i $a_i$ = 0, 1 or 2};\end{array}\right.$ \\ 
$\prod^m_{i,p,k,s=1}L_0(a_i).L_1(a_p).L_2(a_k).L_3(a_s)$ =$\left\{\begin{array}{lll}\mbox{1} & \mbox{\em if $\forall i, p, k, s$ $a_i$ = 0, $a_p$ = 1, $a_k$ = 2, $a_s$ = 3};\\
\mbox{0} & \mbox{\em if $\exists$ i, p, k, s  $a_i$ = 1, 2 or 3, $a_p$ = 0, 2 or 3,};\\ \mbox{ } & \mbox{\em $a_k$ = 0, 1 or 2, $a_s$ = 0, 1 or 2};\end{array}\right.$\\
where $i+p+k+s=m$.

Hence, the minterms for which $f = 1$ are\\
1. $\prod^m_{i=1}L_0(a_i=0)$,
2. $\prod^m_{i=1}L_1(a_i=1)$,
3. $\prod^m_{i=1}L_2(a_i=2)$,\\ 
4. $\prod^m_{i=1}L_3(a_i=3)$,
5. $\prod^m_{i,p,k,s=1}L_0(a_i=0).L_1(a_p=1).L_2(a_k=2).L_3(a_s=3).$

Similarly from Table 2, the minterms for which $f = 2$ are\\  
1. $\prod^m_{i=1}J_0(a_i=0)$,
2. $\prod^m_{i=1}J_1(a_i=1)$,
3. $\prod^m_{i=1}J_2(a_i=2)$,\\ 
4. $\prod^m_{i=1}J_3(a_i=3)$,
5. $\prod^m_{i,p,k,s=1}J_0(a_i=0).J_1(a_p=1).J_2(a_k=2).J_3(a_s=3).$

The minterms for which $f = 3$ are\\  
1. $\prod^m_{i=1}P_0(a_i=0)$,
2. $\prod^m_{i=1}P_1(a_i=1)$,
3. $\prod^m_{i=1}P_2(a_i=2)$, \\
4. $\prod^m_{i=1}P_3(a_i=3)$,
5. $\prod^m_{i,p,k,s=1}P_0(a_i=0).P_1(a_p=1).P_2(a_k=2).P_3(a_s=3)$.
\subsection{Simplification Rules}
Next, we define six simplification rules derived from Table 2\\   
1. $L_i(a).0=0$, $J_i(a).0=0$ and $P_i(a).0=0$\\
2. $L_i(a).1=L_i(a)$, $J_i(a).2=J_i(a)$ and $P_i(a).3=P_i(a)$\\
3. $L_i(a)+0=L_i(a)$, $J_i(a)+0=J_i(a)$ and $P_i(a)+0=P_i(a)$\\
4. $L_i(a)+1=1$, $J_i(a)+2=2$ and $P_i(a)+3=3$\\
5. $L_i(a).L^\prime_{i}(a)=0$, $J_i(a).J^\prime_{i}(a)=0$ and $P_i(a).P^\prime_{i}(a)=0$\\
6. $L_i(a)+L^\prime_{i}(a)=1$, $J_i(a)+J^\prime_{i}(a)=2$ and $P_i(a)+P^\prime_{i}(a)=3$\\
\subsubsection{Simplification Rules for reducing ancilla qudits}
For gate level realization of $L_i$, $J_i$, and $P_i$ we need an ancilla qudit for each of them. Further, to synthesize an $m$-variable quaternary function with $n$ minterms specified in our proposed methodology, we have maximum of $n*m$ ancilla qudits. However, we can reduce the number of ancilla qudits by the following three simplification rules based on the new quaternary $C^2CS$ gate and Table 2:\\

7. $L_i(a)L_j(b)+L_j(a)L_i(b)=C^2CS(a_{ij},b_{ij},0)$, $J_i(a)J_j(b)+J_j(a)J_i(b)=
C^2CS$\\ $(a_{ij},b_{ij},1)$ and $P_i(a)P_j(b)+P_j(a)P_i(b)=C^2CS(a_{ij},b_{ij},2)$ Where $i$, $j$=\{1, 2, 3\}\\

8. $L_i(a).L_i(a)=L_i(a)$, $J_i(a).J_i(a)=J_i(a)$ and $P_i(a).P_i(a)=P_i(a)$\\

9. $L_i(a_1)L_i(a_2)..L_i(a_n)=L_i(a_1,a_2,..,a_n)$,  $J_i(a_1)J_i(a_2)..J_i(a_n)=J_i(a_1,a_2,..,a_n)$, and
$P_i(a_1)P_i(a_2)..P_i(a_n)=P_i(a_1,a_2,..,a_n)$, $i=\{0,1,2,3\}$.\\
\vspace*{-8mm}
\section{Synthesis of Quaternary Functions}
\vspace*{-2mm}
\subsection{$2$-qudit Quaternary Arbitrary Function}
The truth table for the $2$-qudit arbitrary function $f(a, b)$ is given in Table 4.
\begin{table}
\caption {Truth table of $2$-qudit $f(a,b$) }
\begin{center}
\begin{tabular}{|r|r|r|r|r|r|r|r|r|r|r|r|r|r|r|r|r|}
\hline
a , b & 00 & 01 & 02 & 03 & 10 & 11 & 12 & 13 & 20 & 21 & 22 & 23 & 30 & 31 & 32 & 33\\
\hline
$f(a, b)$ & 0 & 3 & 1 & 2 & 3 & 3 & 2 & 0 & 1 & 2 & 1 & 3 & 2 & 1 & 3 & 2 \\
\hline
\end{tabular}
\end{center}
\end{table}

 We re-write the function $f(a, b)$ using our proposed methodology as\\
$f(a, b)$ = $L_0(a)L_2(b)+L_2(a)L_0(b)+L_2(a)L_2(b)+L_3(a)L_1(b)+ J_0(a)J_3(b)+J_1(a)J_2(b)+J_2(a)J_1(b)+J_3(a)J_0(b)++J_3(a)J_3(b) P_0(a)P_1(b)+P_1(a)P_0(b)+P_1(a)P_1(b)+P_2(a)P_3(b)$+$P_3(a)P_2(b)$\\
By using simplification rules $7$ and $9$, we get\\
$f(a, b)$=$C^2CS(a_{02},b_{02},0)+L_2(a, b)+L_3(a)L_1(b)+C^2CS(a_{03},b_{03},1) +C^2CS(a_{12},b_{12},1)+J_3(a,b)+ C^2CS(a_{10},b_{10},2)+C^2CS(a_{23},b_{23},0) +P_1(a,b)$ 
\begin{figure}[!h]
\centering
\vspace*{-5mm}
\includegraphics {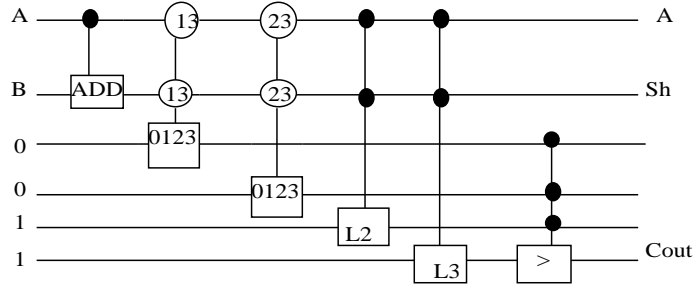}
\vspace*{-3mm}
\caption{Quaternary $2$-qudit Half Adder by our synthesis method}
\label{Figure 1:}
\end{figure}
\vspace*{-10mm}
\subsection{Quaternary Adder}
\vspace*{-3mm}
We synthesize the $2$-qudit half and full adder circuits using our proposed methodology and these circuits are simplified with the use of Modulo4 Addition gate. But the details are not provided due to limitation of space. The gate level implementation of $2$-qudit half and full adder are shown in Figure $8$ and $9$ respectively.
\begin{figure}[!h]
\centering
\vspace*{-5mm}
\includegraphics {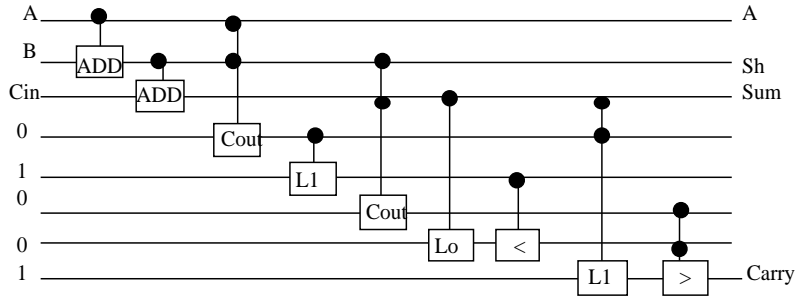}
\vspace*{-5mm}
\caption{Quaternary $2$-qudit Full Adder by our synthesis method}
\label{Figure 1:}
\end{figure}
\vspace*{-5mm}
\subsection{Comparison of Results}
\vspace*{-2mm}
The quantum cost of the half and the full adder circuits, as shown in Figures 10 and 11, and of the benchmark circuits such as $sum2$, $rd53$, $rd73$, $xor5$ are evaluated in terms of the number of M-S gates used. The number of M-S gates required to realize a $2$-qudit Feynman gate, a $3$-qudit Toffoli, $MAX$ and $MIN$ gates are 5, 17, 6, 6 respectively \cite{Chowdhury,Gieseck}. To realize the new GQG, $C^2CS$, ADD gates we need 8 M-S gates for each of them. The M-S gate count cost comparison of our circuits with that  in \cite{Khan1} are given in Table 6. While the second column of the table indicates the maximum number of ancilla qudits required to synthesize the above mentioned circuits, the third column shows the number of ancilla qudits required after using our simplification rules in Section 4.2. The columns 4, 5 and 6 establish that although our design has a small increase in cost for the circuits $rd53$, $rd73$, $xor5$, the results for the $2$-qudit quaternary half and full adder shows that there is more than $50\%$ reduction in the M-S gate count cost compared to \cite{Khan1}.
\vspace*{-3mm}
\begin{table}
\vspace*{-3mm}
\caption{Comparison of Quantum cost by our method vs. \cite{Khan1}}
\vspace*{-5mm}
\begin{center}
\begin{tabular}{|c|c|c|c|c|c|c|}
\hline
& & &\multicolumn{2}{|c|}{\# Levels} &\multicolumn{2}{|c|}{Total Cost}\\
\hline
Circuit & maximum ancilla qudits & Reduced ancilla qudits & Ours & \cite{Khan1} & Ours & \cite{Khan1}\\
\hline
Half Adder & 36 & 6 & 6 & 23 & 46 & 114\\
Full Adder & 120 & 17 & 17 & 40 & 128 & 304\\
Sum2 & 24 & 0 & 4 & - & 8 & -\\
mul2 & 16 & 5 & 5 & - & 40 & -\\
ham3 & 135 & 95 & 25 & - & 135 & - \\
rd53 & 275 & 245 & 15 & - & 120 & -\\
rd73 & 475 & 435 & 35 & - & 280 & -\\
xor5 & 150 & 120 & 7 & - & 56 &  -\\
\hline
\end{tabular}
\end{center}
\end{table}
\vspace*{-9mm}
\section{Conclusion}
\vspace*{-4mm}
In this paper, we have proposed a methodology for logic synthesis of quaternary quantum circuits. We have defined a minterm based approach of expressing a quaternary logic function using $L_i$, $J_i$ and $P_i$ operations. We have also stated the simplification rules for the method.  Quaternary half adder, full adder and some benchmark circuits synthesized by our method, use fewer ternary quantum gates and hence reduce quantum realization cost compared to earlier method in [14]. While the number of levels is fewer in our synthesis, the number of ancilla bits is higher. A synthesis methodology for reduction of the number of ancilla qudits is being investigated.
 



\vspace*{-4mm}
\begin{thebibliography}{1}
\vspace*{-4mm}
\bibitem{Nielsen}
M. A. Nielsen and I. L. Chuang, 
{\em Quantum Computation and Quantum Information,} Cambridge University Press , 2002.
\vspace*{1.1mm}
\bibitem{Ekert}
A. Ekert and  A. Zeilinger, 
{\em The physics of quantum information,} Springer Verlag, Berlin, 2002, pp. 1-14.
\vspace*{1.1mm}
\bibitem{Muthukrishnan}
A. Muthukrishnan and C. R. Stroud Jr., 
{\em Multi-Valued Logic Gates for Quantum Computation,} Phys. Rev. A62, 2000, pp. 052309\/1-8.
\vspace*{1.1mm}
\bibitem{Klimov}
P. Picton, 
{\em A Universal Architecture for Multiple-Valued Reversible Logic,} Multiple-Valued Logic - An International Journal, Vol. 5, 2000, pp. 27-37
\vspace*{1.1mm}
\bibitem{Zilic}
J. L. Brylinski and R. Brylinski, 
{\em Universal Quantum Gates,} (to appear in Mathematics of Quantum Computation, CRC Press, 2002) LANL e-print quant-ph/010862.
\vspace*{-1.1mm}
\bibitem{Yang}
M. Perkowski, A. Al-Rabadi, P. Kerntopf, A. Mishchenko,
and M. Chrzanowska-Jeske, 
{\em Three-Dimensional Realization of Multivalued Functions Using Reversible Logic,} Booklet of 10th Int. Workshop on Post-Binary Ultra-Large-Scale
Integration Systems (ULSI), Warsaw, Poland, May 2001, pp.47- 53
\vspace*{1.1mm}
\bibitem{Chowdhury}
M.M.M Khan, A. K. Biswas, S. Chowdhury, M.Tanzid, K.M.Mohsin, M. Hasan, A. I. Khan , 
{\em Quantum realization of some quaternary circuits,} TENCON 2008 - 2008 IEEE Region 10 Conference , vol., no., pp.1-5, 19-21 Nov. 2008
\vspace*{1.1mm}
\bibitem{Khan}
M. H. Khan,  
{\em Quantum Realization of Quaternary Feynman and Toffoli Gates} Electrical and Computer Engineering, 2006. ICECE '06. International Conference on , vol., no., pp.157-160, 19-21 Dec. 2006
\vspace*{1.1mm}
\bibitem{Mk}
M. H. Khan, 
{\em Synthesis of incompletely specified multi-output quaternary function using quaternary quantum gates,} Computer and information technology, 2007. iccit 2007. 10th international conference on , vol., no., pp.1-6, 27-29 Dec. 2007
\vspace*{1.1mm}
\bibitem{siddi}
M.H.A. Khan, N.K. Siddika, M.A. Perkowski,  
{\em Minimization of Quaternary Galois Field Sum of Products Expression for Multi-Output Quaternary Logic Function Using Quaternary Galois Field Decision Diagram,} Multiple Valued Logic, 2008. ISMVL 2008. 38th International Symposium on , vol., no., pp.125-130, 22-24 May 2008
\vspace*{1.1mm}
\bibitem{SKolay}
S.B.Mandal, A. Chakrabarti, and S. Sur-Kolay, 
{\em Synthesis Technique for Ternary Quantum Logic,} 41 th International Symposium on Multiple-Valued Logic, Tuusula, 2011, pp. 218-223.
\vspace*{1.1mm}
\bibitem{Herman}
R. L. Herrmann, 
{\em Selection and implementation of a ternary switching algebra,} Spring Joint Computer Conference, Atlantic City, 1968, pp. 283-290 .
\vspace*{1.1mm}
\bibitem{Gieseck}
N. Giesecke, D. H. Kim, S. Hossain and M. Perkowski, 
{\em Search for Universal Ternary Quantum Gate Sets with Exact Minimum Costs,} Proceedings of RM Symposium, Oslo, May 16, 2007. 
\vspace*{1.1mm}
\bibitem{Khan1}
M. H. A. Khan, 
{\em A recursive method for synthesizing quantum/reversible quaternary parallel
adder/subtractor with look-ahead carry} Journal of Systems Architecture 54 (2008) 1113–1121
.
\end{thebibliography}
%

\end{document}